\documentclass[useAMS,usenatbib]{mn2e}

\usepackage{graphicx}

\title[The Fundamental Plane for ETGs. Dependence on the magnitude range]
{The Fundamental Plane for early-type galaxies. Dependence on the magnitude range}
\author[A. Nigoche-Netro et al.]{A. Nigoche-Netro$^{1}$\thanks{E-mail:
anigoche@iac.es},
A. Ruelas-Mayorga$^{2}$\thanks{E-mail: rarm@astroscu.unam.mx} 
and A. Franco-Balderas$^{2}$\thanks{E-mail: alfred@astroscu.unam.mx}\\
$^{1}$Instituto de Astrof\'isica de Canarias (IAC), V\'ia L\'actea s/n, 38200 La Laguna, Spain\\
$^{2}$Instituto de Astronom\'ia, Universidad Nacional Aut\'onoma de M\'exico. Apartado Postal 70-264. M\'exico, D.F. M\'exico}

\begin{document}


\pagerange{\pageref{firstpage}--\pageref{lastpage}} \pubyear{2002}

\maketitle

\label{firstpage}

\begin{abstract}

Studying 3 samples of early-type galaxies, which include approximately 8800 galaxies and cover a relatively ample magnitude range ($<\Delta M>$ $\sim 5.5$ $mag$), we find that the coefficients as well as the intrinsic dispersion of the Fundamental Plane depend on the width and brightness of the magnitude range within which the galaxies are distributed. We analyse this dependence and the results show that it is due to the fact that the distribution of galaxies in the space defined by the variables $\log (r_{e}), <\mu>_{e}, \log(\sigma_{0})$ depends on the luminosity.

\end{abstract}


\section{Introduction}

The Fundamental Plane (FP) for ordinary early-type galaxies (ETGs) is a 
correlation between the following observable parameters: {\it
logarithm of the effective radius} ($\log {\kern 1pt} (r_{e} )$),
{\it effective mean} {\it surface brightness} ($<\mu>_{e}$)
and {\it logarithm of the central velocity dispersion} (log $\sigma _{0}$) \citep{djo87,dre87}. This correlation is usually expressed as follows:

\begin{equation}
\log\,(r_{e})\, \; =\, \; a\,\log\,(\sigma_{0})\, \; +\; b\, \, <\mu>_{e}\;
+\; c,
\end{equation}

where $a$, $b$ and $c$ represent scale factors. These scale factors contain important information about ETGs and might depend, among other things, on environment and wavelength.

The FP relation (equation 1) is a consequence of the dynamical equilibrium condition (virial theorem) and of the regular behaviour of the mass-luminosity ratio as well as the structure along the whole range of luminosities for the ETGs. Due to the small intrinsic dispersion ($\sim 0.1$ $dex$ in $r_{e}$ and $\sigma_{0}$, and $\sim 0.1$ in $<\mu>_{e}$), the FP is considered a powerful tool for studying galactic formation and evolution as well as useful for making estimations of large scale distances \citep{kja93,jor96,jor99,kel97}. 

An important projection of the FP is the correlation between log($r_{e}$) and $<\mu>_{e}$, also known as the Kormendy relation (KR) \citep{kor77}:

\begin{equation}
<\mu>_{e}\, \; =\, \; \alpha\; +\; \beta \, \log (r_{e}).
\end{equation}

Different studies have demonstrated that ETGs inside clusters define the KR with an intrinsic dispersion of approximately 0.4 $mag$ in $<\mu>_{e}$ \citep{ham87,hoe87,san91,san01}. Recent studies have demonstrated that ETGs contained in less dense environments and isolated ETGs also follow the KR with the same coefficients and dispersion as do ETGs in clusters \citep{red04,nig07}. 

On the other hand, Nigoche-Netro (2007) and Nigoche-Netro {\it et al}. (2007, 2008) found, using observational data and numerical simulations, that both the coefficients and the intrinsic dispersion for the KR depend on the width of the magnitude range and the brightness of ETGs within the magnitude range. This dependence is caused by a \textquotedblleft geometrical effect\textquotedblright\ due to the fact that the distribution of ETGs on the log($r_{e}$) - $<\mu>_{e}$ 
plane depends on luminosity \citep{var04,don06,nig07} and that the geometric shape of this distribution is not symmetrical. In other words, the geometric shape of the distribution of ETGs on the log($r_{e}$) - $<\mu>_{e}$ plane changes systematically as brighter ETGs are progressively considered, and so the values of the KR coefficients change too, because the fitting of a straight line to a set of data does not give the same result for slope and intercept values for data distributed with a rectangular shape as for data distributed with a triangular shape or, for that matter, with another geometrical shape. 

The previously mentioned results, together with the fact that the KR is a projection of the FP, suggested that a study of the behaviour of the coefficients and intrinsic dispersion of the FP with respect to the magnitude range would prove interesting. Several papers in the literature (Bernardi {\it et al}. 2003c; Jorgensen {\it et al}. 1996) undertake, somewhat differently, this type of investigation. These works have investigated, among other things, the residues of the FP with respect to magnitude. The results they find indicate that there is no correlation between the residues of the FP and magnitude. However, we consider that it is necessary to perform an investigation similar to that presented for the KR in Nigoche-Netro {\it et al}. (2008) for the case of the FP, given that there is no published work in the astronomical literature where this approach is taken.

This study proves to be relevant for investigations that perform comparisons of different galaxy samples such as comparisons of samples for which dependence on the environment, on the redshift or on the wavelength are sought for. If differences are found and the dependence of the coefficients of the FP relation on the magnitude range is not taken into consideration, the differences found may be completely misinterpreted.

In this paper we compile 3 samples of ETGs which include a total of 
approximately 8800 galaxies and cover a relatively ample magnitude range ($<\Delta M>$ $\sim 5.5$ $mag$). It is important to mention that in this work we perform the analysis of the coefficient values and of the intrinsic dispersion of the FP with respect to: 1) the width of the magnitude range (increasing magnitude intervals), that is, we consider the faintest galaxies in each sample and we progressively increase the width of the magnitude interval by inclusion of the brighter galaxies, and 2) the brightness of the magnitude range (narrow magnitude intervals), that is, we consider galaxy samples in progressively brighter fixed-width magnitude intervals. On the other hand, in order to avoid biases in the coefficients, we use the linear regression method known as {\it Measurement errors and Intrinsic Scatter Three dimensional bisector} (MIST$_{Bis}$) \citep{lab00}. This method takes into consideration the errors associated to each one of the variables and minimises the effects of the intrinsic dispersions of each galaxy sample under study. 

In Section 2 we introduce the different galaxy samples. In Section 3 we present the fitting method used for calculating the parameters of the FP and the results from these fits. In this section we also discuss an analysis of the behaviour of the coefficients $a$, $b$ and $c$ and of the intrinsic dispersion for the FP with respect to the absolute magnitude range. Finally, in Section 4, we present our conclusions.

\section{The samples}

{\tiny

\begin{table*}

 \centering
 \begin{minipage}{160mm}
\caption{Name, number of galaxies, magnitude range (according to the papers where the samples were taken from), a
pproximate magnitude range in the B-Filter (calculated by us) and redshift of the different galaxy samples compiled in this work. }

  \begin{tabular}{@{}lccccrr@{}}
\hline

Sample   & N & Magnitude range & Approximate magnitude range in the B-filter & $z$ \\

 \hline

 Total SDSS (g* filter) & 8666  & $-18.0 \ge M_{g*}> -24.1$  & $-17.5 \ge M_{B}> -23.6$ & $\le 0.3$ \\
 Total SDSS (r* filter) & 8666  &  $-18.6 \ge M_{r*}> -24.7$  & $-17.5 \ge M_{B}> -23.6$ & $\le 0.3$ \\
 Total SDSS (i* filter) & 8666  &  $-19.0 \ge M_{i*}> -25.1$  & $-17.5 \ge M_{B}> -23.6$  & $\le 0.3$\\
 Total SDSS (z* filter) & 8666  & $-19.3 \ge M_{z*}> -25.3$  & $-17.5 \ge M_{B}> -23.6$  & $\le 0.3$\\
 Coma cluster (Gunn r filter)& 116 & $-20.4 \ge M_{Gr}> -24.9$  & $-19.3 \ge M_{B}> -23.8$ & $0.024$ \\
 Hydra cluster (Gunn r filter)& 44 & $-19.0 \ge M_{Gr}> -24.6$  & $-17.9 \ge M_{B}> -23.5$  & $0.014$\\

\hline

\end{tabular}
\end{minipage}
\end{table*}

}

We use a sample of 8666 ETGs from the Sloan Digital Sky Survey (total SDSS sample) \citep{ber03}. This sample has surface photometry information in filters g*, r*, i* and z* (absolute magnitude range $-19 \ge M_{r^{*}} > -25$ and its equivalent in other filters). We also use a sample of 116 ETGs for the Coma Cluster \citep{mil97,jor99} in the filter Gunn r ($-20 \ge M_{Gr}> -24.9$) and a 44 ETGs sample in the filter Gunn r ($-19 \ge M_{Gr}> -24$) from the Hydra cluster \citep{mil97}. 

All the samples are redshift-homogeneous (the galaxies are contained within a narrow redshift interval), except for the total SDSS sample which cover a relatively ample redshift interval (0.01 $\leq\;z\;\leq$ 0.3). This sample is magnitude-limited \cite{ber03b}, that is to say that in small volumes and for small values of $z$, it does not contain very bright galaxies, whereas in large volumes and for large $z$ values, it does not contain faint galaxies. Besides, within large volumes there could be evolution effects of the parameters
of the galaxies. So, in order to have a representative sample of the universe in a given volume without any evolution effects it is important to consider narrow redshift intervals. Bernardi {\it et al.} 2003b recommend $\Delta z = 0.04$. This value comes from the sizes 
of the largest structures in the universe seen in numerical simulations of the cold dark matter family of models \citep{col00}. It is also well known that the SDSS photometry underestimates the luminosity of the brightest objects in crowded fields \citep{ber07,ber07b}. In addition, the spectroscopic reduction of the SDSS data underestimates the values of the small central velocity dispersions ($\sigma _{0} \leq\; 150\; km/s$) \citep{ber07b}.

To probe the possible evolution effects and avoid the photometric and spectroscopic biases we have built a subsample from the total SDSS in the redshift interval $0.04\leq\;z\;\leq0.08$ and central velocity dispersion $\sigma _{0} >\; 150\; km/s$. Given that the determination of the distances to the galaxies relies entirely on $z$, we chose the lower limit to keep the influence of the peculiar velocities on the measured $z$ below 10\%. This subsample has 933 galaxies in each filter and cover a magnitude range $<\Delta M>$ $\sim 4$ $mag$ ($-19 \ge M_{g^{*}} > -22.8$ and its equivalent in other filters). From now on, this subsample will be referred to as homogeneous sample from the SDSS.

In Table 1 we present relevant information (number of galaxies, magnitude range and redshift) for the samples of galaxies we use in this paper. The magnitude range information is given in relation to the different filters used (from the literature) and also the approximate range for the B-magnitude (calculated by us). The transformation to the B filter was accomplished by use of the following equation: B - Gunn r = 1.15 \citep{mil97}. The filters g*, r*, i* and z* correspond approximately to the Johnson-Morgan-Cousins filters B, V, R$_{c}$ and I$_{c}$ respectively \citep{fuk96}.

All the samples consider the photometric and spectroscopic parameters log($r_{e}$), $<\mu>_{e}$ and log($\sigma _{0}$) corrected for different biases (galactic extinction, K correction, cosmological dimming and aperture correction). These parameters as well as their uncertainties were taken directly from the different papers cited above. It is worth stressing that the photometric parameters were obtained using de Vaucouleurs $r^{1/4}$ fits. These fits represent a good approximation to the S\'ersic profile-fits because all galaxies used here are bright ($M_{B}\leq -18$) \citep{pru97}. To stress this point we should mention that Nigoche-Netro {\it et al}. (2008) show that the de Vaucouleurs approximation does not produce appreciable biases in the behaviour of the KR parameters with respect to the magnitude range.

Finally, an important characteristic of S0 galaxies is that, in general, the structural properties of their bulges show approximately the same homogeneity as those of E galaxies. Due to this fact, and because the FP uses effective parameters, we use the photometric parameters from the whole galaxy in the case of bright Es, and from the bulge in the case of bright S0s. 


{\tiny

\begin{table*}
 \centering
 \begin{minipage}{125mm}
\caption{FP coefficients for the different samples of galaxies in increasing intervals of magnitude (upper magnitude cut-off).
MI is the absolute magnitude interval within which the galaxies are distributed, $\Delta M$ is the width of the magnitude interval, N is the number of galaxies in the magnitude interval, $a$, $b$ and $c$ are the FP coefficients (obtained using the MIST$_{Bis}$ method) and $\sigma$ is the total intrinsic dispersion of the FP.}


  \begin{tabular}{@{}cccccccrr@{}}
\hline

MI & $\Delta M$ & N   &  $a{}$ & $b$ & $c$   & $\sigma$ \\


\hline

&&&&&&&\\

&&& Total SDSS (g* filter)&&&&\\

\hline

 -18.5 $\geq M$  \textgreater -20.5 &2.0& 1416    &0.949$\pm$0.017& 0.281$\pm$0.027 &-6.864$\pm$0.338&0.272 \\
 -18.5 $\geq M$  \textgreater -21.0 &2.5& 2859    &1.039$\pm$0.008& 0.286$\pm$0.016 &-7.263$\pm$0.177&0.292 \\
 -18.5 $\geq M$  \textgreater -21.5 &3.0& 4760   &1.150$\pm$0.005& 0.293$\pm$0.011&-7.718$\pm$0.136&0.302  \\
 -18.5 $\geq M$  \textgreater -22.0 &3.5& 6693   &1.266$\pm$0.006& 0.301$\pm$0.009&-8.136$\pm$0.146&0.310  \\
 -18.5 $\geq M$  \textgreater -22.5 &4.0& 7997   &1.349$\pm$0.006& 0.307$\pm$0.007&-8.444$\pm$0.158&0.314  \\
 -18.5 $\geq M$  \textgreater -23.0 &4.5& 8517    &1.393$\pm$0.007&0.309$\pm$0.007 &-8.596$\pm$0.166&0.317  \\
 -18.5 $\geq M$  \textgreater -23.5 &5.0& 8645    &1.408$\pm$0.007&0.311$\pm$0.007 &-8.670$\pm$0.171&0.317  \\
 -18.5 $\geq M$  \textgreater -24.0 &5.5& 8661    &1.411$\pm$0.007& 0.312$\pm$0.007&-8.688$\pm$0.172&0.317  \\

&&&&&&&\\

&&&Total SDSS (r* filter)&&&&\\

\hline

 -19.0 $\geq M$  \textgreater -21.0 &2.0& 1010    &0.994$\pm$0.022& 0.297$\pm$0.035 &-6.917$\pm$0.444&0.267  \\
 -19.0 $\geq M$  \textgreater -21.5 &2.5& 2243    &1.062$\pm$0.010& 0.296$\pm$0.019 &-7.184$\pm$0.221&0.282  \\
 -19.0 $\geq M$  \textgreater -22.0 &3.0& 4051   &1.158$\pm$0.006& 0.298$\pm$0.012&-7.535$\pm$0.145&0.291   \\
 -19.0 $\geq M$  \textgreater -22.5 &3.5& 6053   &1.271$\pm$0.006& 0.305$\pm$0.009&-7.956$\pm$0.145&0.295   \\
 -19.0 $\geq M$  \textgreater -23.0 &4.0& 7649    &1.359$\pm$0.007&0.311$\pm$0.007 &-8.265$\pm$0.154&0.297   \\
 -19.0 $\geq M$  \textgreater -23.5 &4.5& 8406    &1.406$\pm$0.007&0.312$\pm$0.007 &-8.408$\pm$0.160&0.297   \\
 -19.0 $\geq M$  \textgreater -24.0 &5.0& 8627    &1.425$\pm$0.007& 0.315$\pm$0.007&-8.495$\pm$0.165&0.298   \\
 -19.0 $\geq M$  \textgreater -24.7 &5.7& 8663    &1.428$\pm$0.008& 0.315$\pm$0.007&-8.516$\pm$0.167&0.298   \\

&&&&&&&\\

&&&Total SDSS (i* filter)&&&&\\

\hline

 -19.0 $\geq M$  \textgreater -21.0 &2.0& 506    &1.001$\pm$0.033& 0.299$\pm$0.053 &-6.817$\pm$0.654&0.265  \\
 -19.0 $\geq M$  \textgreater -21.5 &2.5& 1363    &1.048$\pm$0.016& 0.300$\pm$0.028 &-7.045$\pm$0.331&0.270  \\
 -19.0 $\geq M$  \textgreater -22.0 &3.0& 2809   &1.122$\pm$0.008& 0.302$\pm$0.016&-7.361$\pm$0.182&0.279  \\
 -19.0 $\geq M$  \textgreater -22.5 &3.5& 4714   &1.217$\pm$0.006& 0.307$\pm$0.011&-7.723$\pm$0.145&0.281  \\
 -19.0 $\geq M$  \textgreater -23.0 &4.0& 6648    &1.321$\pm$0.007&0.314$\pm$0.008 &-8.112$\pm$0.151&0.286  \\
 -19.0 $\geq M$  \textgreater -23.5 &4.5& 7988    &1.386$\pm$0.007&0.317$\pm$0.007 &-8.317$\pm$0.156&0.287  \\
 -19.0 $\geq M$  \textgreater -24.0 &5.0& 8514    &1.422$\pm$0.008& 0.319$\pm$0.007&-8.437$\pm$0.162&0.287  \\
 -19.0 $\geq M$  \textgreater -25.0 &6.0& 8664    &1.434$\pm$0.008& 0.320$\pm$0.007&-8.493$\pm$0.167&0.287  \\

&&&&&&&\\

&&&Total SDSS (z* filter)&&&&\\

\hline

 -19.5 $\geq M$  \textgreater -22.0 &2.5& 1802    &1.129$\pm$0.013& 0.306$\pm$0.022 &-7.236$\pm$0.285&0.279  \\
 -19.5 $\geq M$  \textgreater -22.5 &3.0& 3460    &1.192$\pm$0.008& 0.305$\pm$0.013 &-7.492$\pm$0.171&0.279  \\
 -19.5 $\geq M$  \textgreater -23.0 &3.5& 5448   &1.280$\pm$0.007& 0.310$\pm$0.009&-7.833$\pm$0.151&0.280   \\
 -19.5 $\geq M$  \textgreater -23.5 &4.0& 7238   &1.346$\pm$0.007& 0.316$\pm$0.008&-8.097$\pm$0.149&0.280   \\
 -19.5 $\geq M$  \textgreater -24.0 &4.5& 8220    &1.387$\pm$0.007&0.319$\pm$0.007 &-8.245$\pm$0.151&0.281   \\
 -19.5 $\geq M$  \textgreater -24.5 &5.0& 8591    &1.400$\pm$0.007&0.320$\pm$0.007 &-8.307$\pm$0.153&0.282   \\
 -19.5 $\geq M$  \textgreater -25.0 &5.5& 8658    &1.405$\pm$0.003& 0.321$\pm$0.007&-8.331$\pm$0.155&0.284   \\
 -19.5 $\geq M$  \textgreater -25.4 &5.9& 8665    &1.406$\pm$0.007& 0.321$\pm$0.007&-8.338$\pm$0.156&0.284   \\

&&&&&&&\\

&&&Coma (Gunn r filter)&&&&\\

\hline

 -20.4 $\geq M$  \textgreater -22.5 &2.1& 101    &0.888$\pm$0.182&0.294$\pm$0.135 &-7.102$\pm$3.350&0.361   \\
 -20.4 $\geq M$  \textgreater -23.0 &2.6& 110   &0.974$\pm$0.122&0.297$\pm$0.111 &-7.365$\pm$2.220&0.390   \\
 -20.4 $\geq M$  \textgreater -23.5 &3.1& 113   &1.002$\pm$0.093&0.299$\pm$0.099 &-7.522$\pm$1.698&0.391   \\
 -20.4 $\geq M$  \textgreater -24.0 &3.6& 114    &1.018$\pm$0.080&0.301$\pm$0.097 &-7.633$\pm$1.460&0.391  \\
 -20.4 $\geq M$  \textgreater -24.9 &4.5& 116    &1.068$\pm$0.052&0.309$\pm$0.096 &-7.943$\pm$1.014&0.393  \\

&&&&&&&\\

&&&Hydra (Gunn r filter)&&&&\\

\hline

 -19.0 $\geq M$  \textgreater -21.5 &2.5& 28    &1.337$\pm$0.099&0.307$\pm$0.116 &-8.390$\pm$2.295&0.281    \\
 -19.0 $\geq M$  \textgreater -22.0 &3.0& 35    &1.424$\pm$0.097&0.305$\pm$0.086 &-8.606$\pm$2.180&0.273    \\
 -19.0 $\geq M$  \textgreater -22.5 &3.5& 40    &1.506$\pm$0.105&0.307$\pm$0.075 &-8.797$\pm$2.288&0.295    \\
 -19.0 $\geq M$  \textgreater -23.5 &4.5& 43   &1.532$\pm$0.116&0.308$\pm$0.074 &-8.901$\pm$2.509&0.286    \\
 -19.0 $\geq M$  \textgreater -24.6 &5.1& 44   &1.596$\pm$0.151&0.317$\pm$0.087 &-9.224$\pm$3.202&0.287    \\

\hline

\end{tabular}
\end{minipage}
\end{table*}
 }

\section[]{The FP relation for the different samples of ETGs}

\subsection{Calculation of the FP relation}

The estimation of the FP coefficients may be severely affected by the fitting method used and by the independent variable adopted for the fit. The biases may be larger if there are measurement errors in the variable, if these errors are correlated or if there is intrinsic dispersion. The {\it Measurement errors and Intrinsic Scatter Three dimensional bisector} method (MIST$_{Bis}$) \citep{lab00} is a statistical model that takes into consideration the different sources of bias mentioned above. Here we use this method to calculate the FP coefficients. This fitting method is a 3 dimensional generalisation of the {\it Bivariate Correlated Errors and Intrinsic Scatter bisector} ($BCES_{Bis}$) method described by Isobe {\it et al}. (1990) and Akritas \& Bershady (1996). The coefficients for the edge-on FP are calculated using the BCES$_{Bis}$ method.

From the photometric and spectroscopic parameters for the different galaxy samples we calculate the FP coefficients in increasing and  narrow magnitude intervals (see Section 1), the errors of these coefficients and the total intrinsic dispersion ($\sigma$) of the FP relation (subtracting in quadrature from the intrinsic dispersion in $<\mu>_{e}$ the residues dispersion due to the measurement errors of $<\mu>_{e}$, log($r_{e}$) and log($\sigma _{0}$)). The definition of the total intrinsic dispersion of the FP is a generalisation in three dimensions of the definition of the total intrinsic dispersion of the KR given by La Barbera {\it et al.} (2003). According to La Barbera {\it et al.} (2003), for the calculation of the total intrinsic dispersion it is necessary to have the measurement errors of each one of the variables involved. These errors were taken from the literature (see Section 2.1), while the theoretical errors in the FP coefficients were calculated by us following La Barbera {\it et al.} (2000) in 1$\sigma$ intervals. 

In Tables 2 and 4 we present the results for the values of the
coefficients of the FP for the different samples of galaxies and for
different magnitude intervals.


\begin{figure*}

\includegraphics[angle=-90,width=14cm]{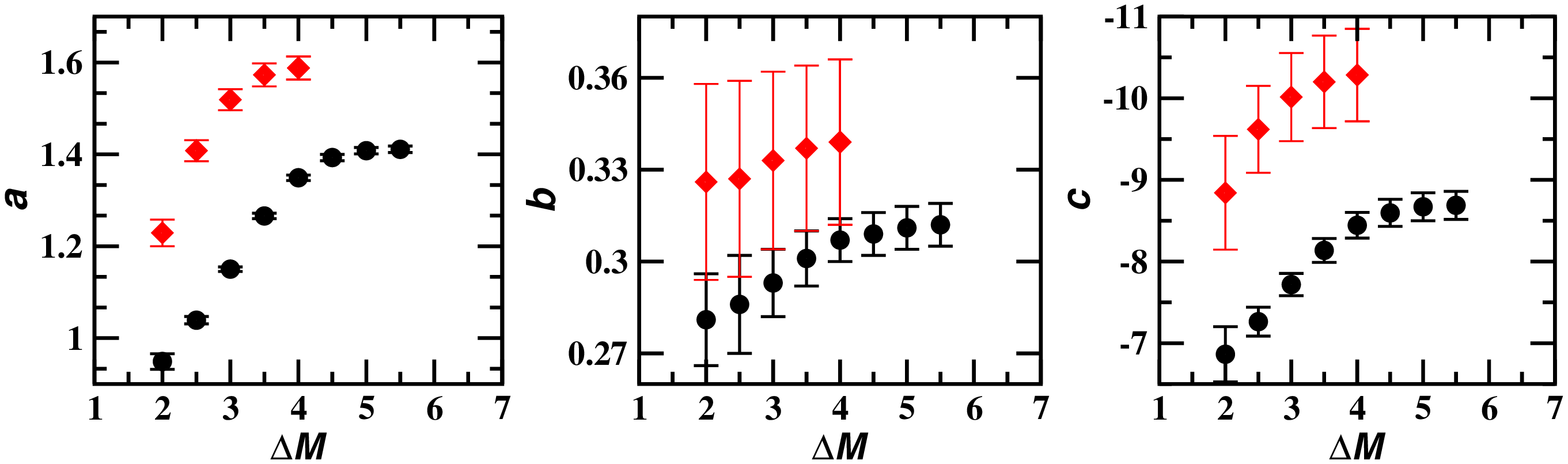}

  \caption{Behaviour of the FP coefficients with respect to the width 
   of the magnitude range (upper magnitude cut-off). Circles represent the total SDSS sample (g* filter). Diamonds represent the homogeneous SDSS sample (g* filter).}

\end{figure*}


\subsection{Behaviour of the FP coefficients with respect to the width of the magnitude range}

As may be seen from the results given in Table 2 for the different galaxy samples, the intrinsic dispersion increases when a growing number of brighter galaxies is included in the samples (upper magnitude cut-off), that is to say when we use increasing magnitude intervals. We also observe that the $a$, $b$ and $c$ coefficients increase systematically as brighter galaxies are included (see Fig. 1) and that these changes are larger than the associated errors for most of the cases (differences between $a$ coefficients may be as large as 33\%). On the other hand, if we consider the brightest galaxies in each sample and we progressively increase the width of the magnitude interval by inclusion of the fainter galaxies (lower magnitude cut-off), the behaviour of the intrinsic dispersion and FP coefficients is similar to the one described above, that is, the FP parameters increase systematically when we increase the width of the magnitude interval.

To investigate whether the changes in the coefficients are significant or whether they are only the product of statistical fluctuations, it is necessary to apply non-parametric tests to the data for the different galaxy samples. Among the non-parametric tests appropriate for our study we find the run test \citep{ben66,nig308}. 

In Table 3 we show the results of the application of the run test method to the data for coefficients $a$, $b$ and $c$. We take as a null hypothesis that there is no underlying trend in the data. The percentages given in Table 3 refer to the confidence level with which we may reject the null hypothesis. From Table 3 we may see that, on average, the null hypothesis may be rejected with a confidence level of 96\%. This implies that there are strong reasons to affirm that there is an underlying trend in the values of each one of the coefficients of the FP. 


In appendix A we present an analysis of the robustness of the run test using the bootstrap methodology \citep{efr79}. This analysis includes measurement errors associated with the data. The conclusions of this analysis indicate that the run test is very efficient in finding underlying trends in sets of data even if the errors associated are quite large.


Here, it is important to mention that when we compare the behaviour of the FP coefficients for the total SDSS sample and the homogeneous SDSS sample (Fig. 1) we note that this behaviour is similar for both samples, meaning that the coefficients for the homogeneous sample increase systematically as we increase the width of the magnitude range. It is also clear that the values of these coefficients result in larger values than those for the total sample. However, it is not possible to say that this diference may be caused by evolution, because the biases in the structural parameters, and the cut in the velocity dispersion might also have an effect on the values of the coefficients (see Section 3.5 for full details).

{\tiny

\begin{table}


 \centering
 \begin{minipage}{62mm}

\caption{Run test for the evaluation of the FP coefficients (increasing magnitude intervals) from the different samples of galaxies.
The null hypothesis establishes that there is no underlying trend in the data, the percentages refer to the confidence level with 
which we can reject the null hypothesis. }

  \begin{tabular}{@{}lccccr@{}}
  
\hline

Sample   & $a$ & $b$ & $c$ \\

 \hline

 Total SDSS (g* filter) & 98\% & 98\% & 98\% \\
 Total SDSS (r* filter) & 98\% & 98\% & 98\% \\
 Total SDSS (i* filter) & 98\% & 98\% & 98\% \\
 Total SDSS (z* filter) & 98\% & 98\% & 98\% \\
 Coma (Gunn r filter)& 93\% & 93\% & 93\%  \\
 Hydra (Gunn r filter)& 93\% & 93\% & 93\%  \\

\hline

\end{tabular}
\end{minipage}

\end{table}


}

\subsection{Behaviour of the FP coefficients with respect to the brightness of the magnitude range}

In order to analyse the behaviour of the FP coefficients with respect to the brightness of the magnitude range and study the distribution of the galaxies in the space defined by the $\log (r_{e}), <\mu>_{e}, \log(\sigma)$ variables, we may utilise the FP edge-on graphs. Figure 2  shows the results for the total SDSS sample in the g* filter, where each symbol and colour represent a one-magnitude wide interval and the thick continuous line represents the fit to all the data. Table 4 shows the values of the FP coefficients. These values were obtained from the fits ($BCES_{Bis}$) made to the different magnitude intervals, however in order to analyse the effects that the different independent variables may have on the results of the fits, we present the results using as independent variables $\log (r_{e})$ and $<\mu>_{e}$.


\begin{figure*}

\centering
\includegraphics[bb= 20 20 750 750,angle=-90,width=12cm,clip]
{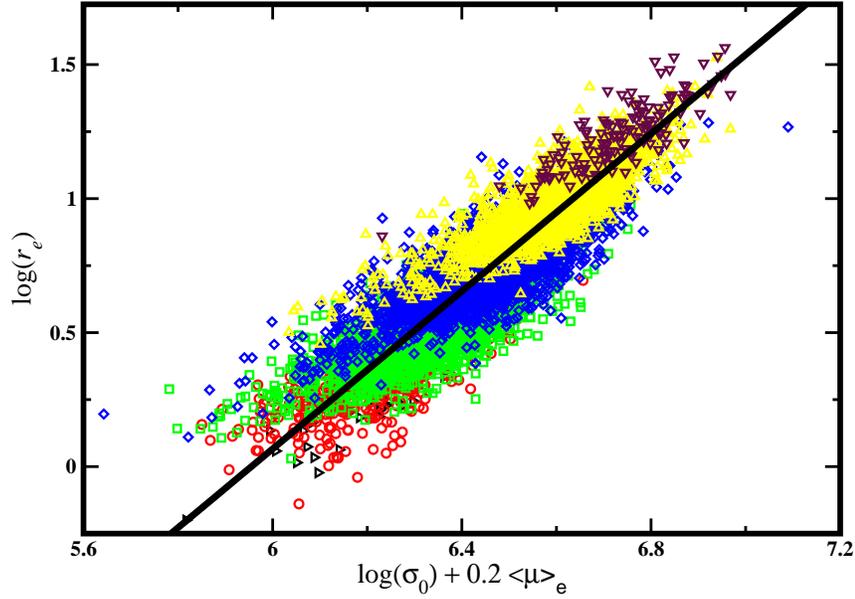}

 \vspace*{-70pt}

  \caption{Distribution of the galaxies in the complete SDSS sample (g* filter) on the edge-on plane. Each symbol and colour represent a one-magnitude wide interval. The continuous line is the fit (BCES$_{Bis}$) obtained for all the data in the sample.}

\end{figure*}



{\tiny

\begin{table*}
 \centering
 \begin{minipage}{125mm}
\caption{FP coefficients for the total SDSS sample (g* filter) in narrow magnitude intervals. MI is the absolute magnitude interval within as the galaxies are distributed, $\Delta M$ is the width of the magnitude interval, N is the number of galaxies in the magnitude interval, $a$, $b$ and $c$ are the FP coefficients and 
$\sigma$ is the intrinsic dispersion of the FP.}

  \begin{tabular}{@{}cccccccrr@{}}
\hline

MI & $\Delta M$ & N   &  $a{}$ & $b$ & $c$   & $\sigma$ \\

\hline

&&&&&&&\\

&&&BCES$_{Bis}$ &&&&\\

\hline

 -18.5 $\geq M$  \textgreater -19.0 &0.5& 26    &1.145$\pm$0.050& 0.229$\pm$0.002&-6.874$\pm$0.315&0.240    \\
 -19.0 $\geq M$  \textgreater -20.0 &1.0& 492    &1.157$\pm$0.029& 0.231$\pm$0.001 &-6.862$\pm$0.185&0.294 \\
 -20.0 $\geq M$  \textgreater -21.0 &1.0& 2341    &1.126$\pm$0.012& 0.225$\pm$0.001 &-6.603$\pm$0.075&0.289 \\
 -21.0 $\geq M$  \textgreater -22.0 &1.0& 3834   &1.159$\pm$0.009& 0.232$\pm$0.001&-6.738$\pm$0.195&0.274  \\
 -22.0 $\geq M$  \textgreater -23.0 &1.0& 1824   &1.220$\pm$0.012& 0.244$\pm$0.001&-7.062$\pm$0.081&0.259  \\
 -23.0 $\geq M$  \textgreater -24.0 &1.0& 144   &1.330$\pm$0.055& 0.266$\pm$0.001&-7.719$\pm$0.371&0.228  \\

&&&&&&&\\

&&& BCES$_{log(r_{e})}$ &&&&\\

\hline

 -18.5 $\geq M$  \textgreater -19.0 &0.5& 26    &1.267$\pm$0.111& 0.253$\pm$0.001&-7.630$\pm$0.691&0.062    \\
 -19.0 $\geq M$  \textgreater -20.0 &1.0& 492    &1.591$\pm$0.079& 0.318$\pm$0.001 &-9.562$\pm$0.492&0.083 \\
 -20.0 $\geq M$  \textgreater -21.0 &1.0& 2341    &1.504$\pm$0.031& 0.301$\pm$0.001 &-9.000$\pm$0.199&0.083 \\
 -21.0 $\geq M$  \textgreater -22.0 &1.0& 3834   &1.458$\pm$0.024& 0.292$\pm$0.001&-8.673$\pm$0.158&0.075  \\
 -22.0 $\geq M$  \textgreater -23.0 &1.0& 1824   &1.448$\pm$0.032& 0.290$\pm$0.001&-8.569$\pm$0.212&0.069  \\
 -23.0 $\geq M$  \textgreater -24.0 &1.0& 144   &1.480$\pm$0.122& 0.296$\pm$0.001&-8.726$\pm$0.826&0.062  \\

&&&&&&&\\

&&& BCES$_{<\mu>_{e}}$ &&&&\\

\hline

 -18.5 $\geq M$  \textgreater -19.0 &0.5& 26    &1.492$\pm$0.019& 0.244$\pm$0.016&-6.900$\pm$0.326&0.312    \\
 -19.0 $\geq M$  \textgreater -20.0 &1.0& 492    &1.771$\pm$0.090& 0.266$\pm$0.001 &-7.245$\pm$0.141&0.435 \\
 -20.0 $\geq M$  \textgreater -21.0 &1.0& 2341    &1.731$\pm$0.044& 0.263$\pm$0.001 &-7.107$\pm$0.069&0.429 \\
 -21.0 $\geq M$  \textgreater -22.0 &1.0& 3834   &1.630$\pm$0.036& 0.255$\pm$0.001&-6.864$\pm$0.059&0.395  \\
 -22.0 $\geq M$  \textgreater -23.0 &1.0& 1824   &1.586$\pm$0.048& 0.252$\pm$0.001&-6.707$\pm$0.081&0.363  \\
 -23.0 $\geq M$  \textgreater -24.0 &1.0& 144   &1.613$\pm$0.163& 0.254$\pm$0.013&-6.659$\pm$0.276&0.322  \\

\hline

\end{tabular}
\end{minipage}
\end{table*}
 }

In Figure 2 and Table 4 we can see that different magnitude galaxies are progressively displaced from the lower part of the diagram (faint galaxies) towards the upper part (bright galaxies) and that the values of the coefficients and of the dispersion of the FP change systematically. These changes are larger than the associated errors for most of the cases (application of the run test confirms that there exists an underlying trend in the data for each one of the coefficients). It is also clear that the fit to the complete sample shows significant differences with the fits performed to galaxies of different luminosity. For example, the differences between coefficients $a$, $b$ and $c$ for the middle part of the galaxy distribution (-21.0 $\geq M$  \textgreater -22.0) and the $a$, $b$ and $c$ coefficients for the total sample (-18.5 $\geq M$  \textgreater -24.0) is approximately 20\%. From what we have just said it can be concluded that different luminosity galaxies define consecutive planes with different values for the coefficients and the dispersion. This implies that the distribution of galaxies in the space defined by variables $\log (r_{e}), <\mu>_{e}, \log(\sigma)$ changes systematically as brighter galaxies are considered (see Section 3.5 for full details).

What we have just described goes in the same direction as what Bender {\it et al.} (1992) reported which indicates that the distribution of different stellar systems (elliptical, spiral bulges, compact elliptical and dwarf ellipticals) in the parameter space $\log (r_{e}), <\mu>_{e}, \log(\sigma)$ depends on luminosity. According to Bender {\it et al.} (1992), this means that the different systems define planes on this space with possible small, parallel offsets between systems.

An alternative way for studying the behaviour of the FP as a function of the magnitude range, is to analyse the values and distribution of the residues, that is to say, to analyse the difference between the predicted values from the best fit to the FP data and the real values. However, given that the values of the coefficients of the FP for the entire sample are different from the values of the coefficients of the samples in different magnitude ranges, we shall analyse the behaviour of the residues taking the coefficients both for the entire sample (-18.5 $\geq M$  \textgreater -24.0) as well as for the middle part of the distribution (-21.0 $\geq M$  \textgreater -22.0 ). In Figures 3 and 4 we show the orthogonal residues of the FP (g* filter) with respect to the $\log (r_{e}), <\mu>_{e}, \log(\sigma)$, $M$ and $X_{FP}$ variables. Where $X_{FP}$ represents the distance along the major axis of the FP (see Bernardi {\it et al}. 2003c).

As can be seen from Figure 3, when the residues are obtained considering the coefficients of the FP for the complete sample, the distribution of the residues with respect to the different variables which we have used show no correlation except, perhaps, for the case of the $\log(\sigma)$ and $X_{FP}$ variable. However, when we consider the FP coefficients for the middle part of the galaxy distribution (-21.0 $\geq M$  \textgreater -22.0)(Figure 4) we find that the residue distribution with respect to the $\log (r_{e})$, $M$ and $X_{FP}$ variables shows a correlation. In particular for the magnitude case the correlation is very clear. Therefore, the residue distribution confirms the fact that different luminosity galaxies are found in consecutive planes characterised by different coefficients. But why do the values of the coefficients of the FP for different luminosity galaxies change? Section 3.5 will deal in greater detail with this.


\begin{figure*}

\centering
\includegraphics[bb= 20 20 750 750,angle=-90,width=12cm,clip]
{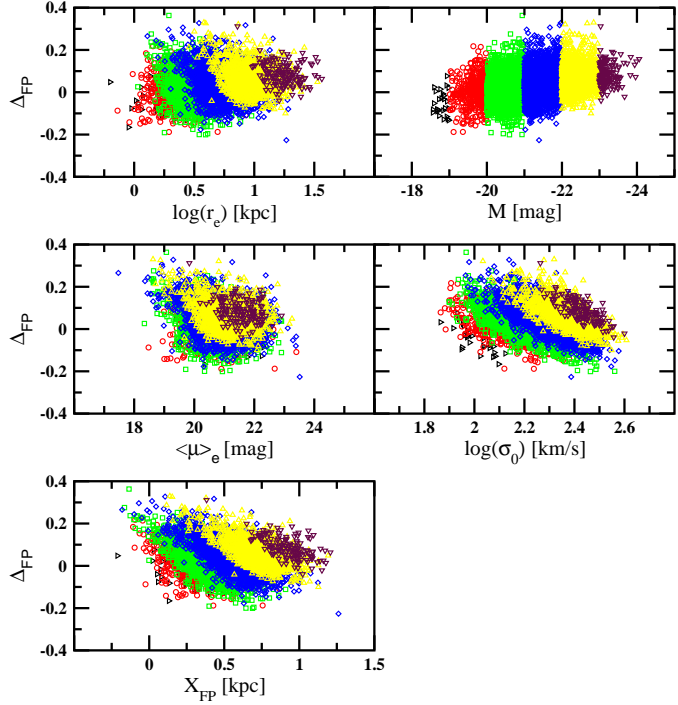}

 \vspace*{-70pt}

  \caption{Orthogonal residues for the FP vs. $\log (r_{e}), M, <\mu>_{e}, \log(\sigma)$, and $X_{FP}$. The FP coefficients we use correspond to the $BCES_{Bis}$ fit for all the galaxies in the SDSS sample (g* filter). Each symbol and colour represent a one-magnitude wide interval.}

\end{figure*}



\begin{figure*}

\centering
\includegraphics[bb= 20 20 750 750,angle=-90,width=12cm,clip]
{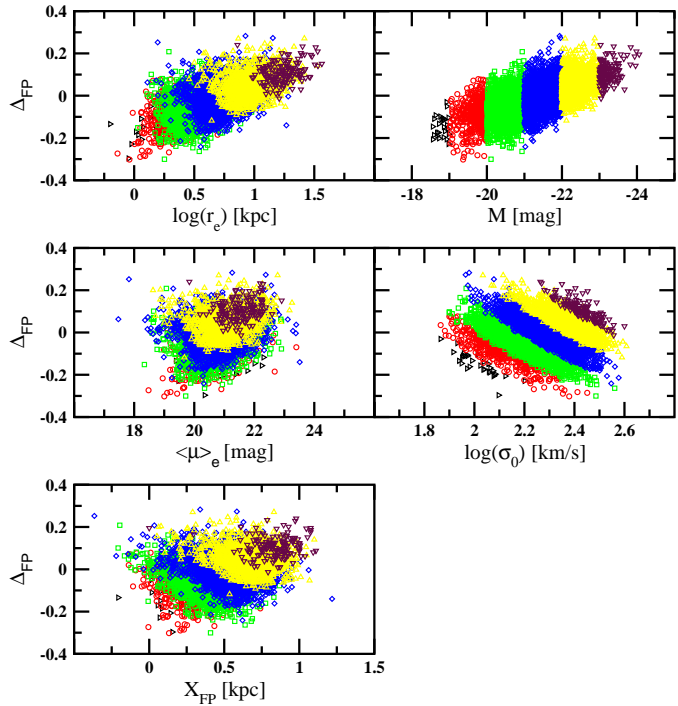}

 \vspace*{-70pt}

  \caption{Orthogonal residues for the FP vs. $\log (r_{e}), M, <\mu>_{e}, \log(\sigma)$, and $X_{FP}$. The FP coefficients we use correspond to the  $BCES_{Bis}$ fit for the middle part (-21.0 $\geq M$  \textgreater -22.0) of the SDSS galaxy distribution (g* filter). Each symbol and colour represent a one-magnitude wide interval.}

\end{figure*}


\subsection{Previous works dealing with the behaviour of the FP with respect to magnitude}


\begin{figure*}

\centering
\includegraphics[bb= 20 20 750 750,angle=-90,width=12cm,clip]
{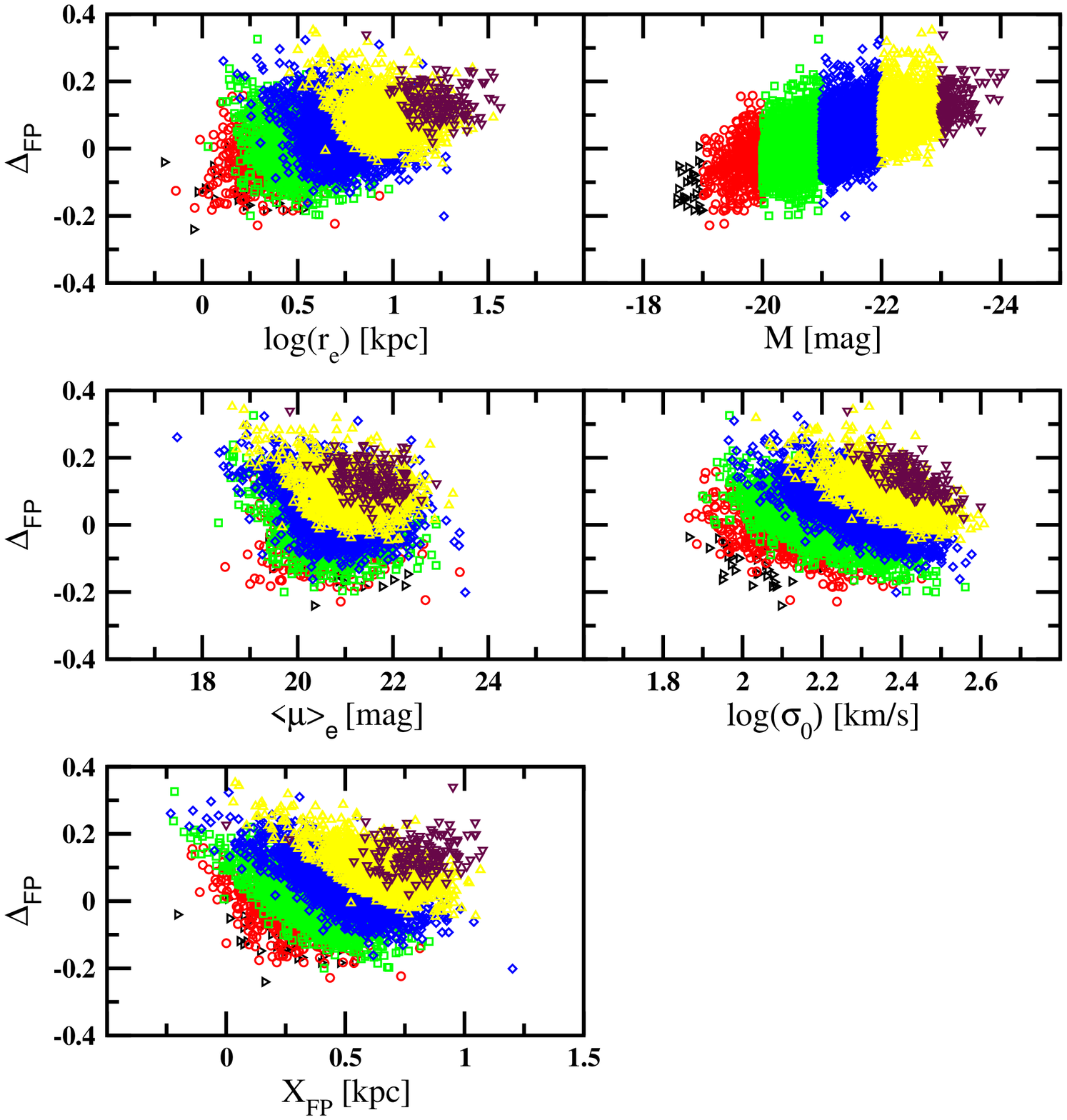}

 \vspace*{-70pt}

  \caption{Orthogonal residues for the FP vs. $\log (r_{e}), M, <\mu>_{e}, \log(\sigma)$, and $X_{FP}$. The FP coefficients correspond to the direct fit ($\log (r_{e})$ independent variable) for Table 3 from Bernardi {\it et al}. (2003c)(complete sample of the SDSS in filter g*). Each symbol and colour represent a one-magnitude wide interval. }

\end{figure*}


In the literature there are several papers which study the dependence of the FP with magnitude, for example Jorgensen {\it et al}. (1996) and Bernardi {\it et al}. (2003c); utilising the residues of the FP they find that there is no dependence of the FP with magnitude. However, the type of analysis undertaken by them might be masking the dependence on the magnitude range which we are reporting in this paper. Given that in the Bernardi {\it et al}. (2003c) paper they use the exact same sample of ETGs from the SDSS as we do here, in what follows we shall analyse in more detail their results.

Among the most important conclusions of Bernardi {\it et al}. (2003c) related to our work, we find that the orthogonal residues of the FP (when they take as reference the orthogonal fit coefficients) with respect to the $\log (r_{e}), <\mu>_{e}, \log(\sigma)$, $M$ and $X_{FP}$ variables do not show an appreciable correlation, except for the $\log(\sigma)$ and $X_{FP}$ cases. On the other hand, when they analyse the residues taking the direct fit FP coefficients and using as independent variable $\log (r_{e})$, they mention that no correlations are found. They interpret the correlation found for the orthogonal fit as a projection effect which affects this type (orthogonal) of fits but not the direct fits (if $\log (r_{e})$ is taken as dependent variable). However in figure 5 we show the results of the orthogonal residues (g* filter) taking the data from the direct fit ($\log (r_{e})$ is the independent variable) from the table 3 of Bernardi {\it et al}. (2003c) and it can be clearly noticed that there is a correlation of the residues with respect to the $\log (r_{e})$ and magnitude. Particularly for the magnitude case the correlation is perfectly clear and apparent. On the other hand, the data from table 4 show that the change in the coefficients of the FP occurs no matter which variable is utilised as independent variable in the fitting process. So we are able to ensure that the fitting method and the independent variable utilised for the fits are not responsible for the dependence of the FP on the magnitude range which we are reporting in this paper.


The distribution of the residues for the different variables presented in figure 3 has been obtained in a similar manner as that followed by Bernardi {\it et al}. (2003c), that is to say, taking as reference the FP coefficients for the total sample and taking into consideration the different sources of bias. When we compare these distributions with those presented in Figs. 2 and 3 of Bernardi {\it et al}. (2003c), we notice that the residues distribution is similar, that is, there are no appreciable correlations between the values of the residues and the different variables which we analyse (except for the $\log(\sigma)$ and $X_{FP}$ cases). However, as we mention in Section 3.3, when we take as reference a representative plane of the sample in one-magnitude wide intervals, for example the plane at the middle part of the distribution of galaxies (-21.0 $\geq M$  \textgreater -22.0) (see figure 4), the residues clearly show a correlation with respect to the $\log (r_{e})$, $M$ and $X_{FP}$ variables, in particular, the residues show a clear correlation with respect to magnitude. From all of this we can say with certainty that in the Bernardi {\it et al}. (2003c) analysis the dependence of the FP with the range of magnitude has been masked due to the way in which these authors performed the analysis.

\subsection{Comparison of the behaviour of the FP and the KR with respect to the magnitude range}

Nigoche-Netro (2007) and Nigoche-Netro {\it et al}. (2007, 2008)
demonstrated, both with observational data as well as with 
numerical simulations that the coefficients and the intrinsic 
dispersion of the KR depend on the width and brightness of the magnitude range. This dependence is caused by a ``geometrical effect" due to the fact that the distribution of the ETGs on the log($r_{e}$) - $<\mu>_{e}$ plane depends on luminosity (different brightness ETGs are distributed in parallel lines on the log($r_{e}$) - $<\mu>_{e}$ plane) and that the geometric shape of the distribution of the ETGs on this plane is not symmetrical. Given this fact, and since the KR is a projection of the FP, the dependence of the values of the FP coefficients on the width and brightness of the magnitude range might be caused by a geometrical effect due to the form of the distribution of the ETGs on the space defined by the variables $\log (r_{e}), <\mu>_{e}, \log(\sigma)$ and not by intrinsic physical properties of the ETGs. In fact this is the case given that in Section 3.3 we have seen that the distribution of ETGs in the $\log (r_{e}), <\mu>_{e}, \log(\sigma)$ space depends on luminosity, besides in Section 3.2
 and 3.3 we have established that the values of the coefficients and of the dispersion of the FP change systematically as we consider brighter galaxies. This implies that the geometric shape of the galaxy distribution changes as brighter galaxies are included. This may easily be checked from figure 3 where one can see that as brighter galaxies are considered the shape of the galaxy distribution changes systematically.

From the aforementioned, we may infer that the FP coefficients and its intrinsic dispersion depend on the width and brightness of the magnitude range. This dependence is caused by a geometrical effect due to the fact that the distribution of the ETGs on the $\log (r_{e}), <\mu>_{e}, \log(\sigma)$ space depends on luminosity, in other words, the geometric form of the distribution of ETGs on the $\log (r_{e}), <\mu>_{e}, \log(\sigma)$ space changes systematically as we consider brighter ETGs, and so the values of the FP coefficients change too, because the fitting of a plane to a set of data does not give the same result for data distributed with a cubic form as for data distributed with a piramidal form or, for that matter, with another geometrical form. In this sense, any other systematic restrictions imposed on a sample of ETGs, such as brightness cuts, effective radius cuts or velocity dispersion cuts will cause changes in the geometric form of the distribution of ETGs. The more pronounced these changes are made, the more pronounced will be the changes in the values of the FP coefficients. This could be corroborated with the data in figure 1, where we show that the coefficients of the FP for the homogeneous sample are larger than those for the complete sample. This behaviour is due to the cut in $\log(\sigma)$ and not to the fact that the FP for this galaxy subsample might be intrinsically different.

In view of what we present above, the traditional FP relation (equation 1) could be written as follows:

\begin{equation}
\log\,(r_{e})\, \; =\, \; a'\,\log\,(\sigma_{0})\, \; +\; b'\, \, <\mu>_{e}\;
+\; c'
\end{equation}

where $a^{\prime}$, $b^{\prime}$ and $c^{\prime}$ depend on the width and brightness of the magnitude range. 

It is very important to be careful with equations 1 and 3 because, in order to be consistent, the first one should cover the whole range of magnitudes and the second should correspond to a particular magnitude range. In other words, the traditional FP relation will only produce the real coefficients when complete samples, as far as magnitude is concerned, are considered. 

Here it should be mentioned that in the literature there are extensive studies of the FP for different galaxy samples. The values of the coefficients of the FP which are reported therein show important differences among the various authors. For example D'Onofrio {\it et al}. (2006) have made a study of several papers that have dealt with the FP. From the D'Onofrio {\it et al}. (2006) data we find that the maximum difference of the values of the $a$ coefficient estimated by different authors is of the order 31\%, for the $b$ coefficient the maximum difference amounts to 16\% and for the $c$ coefficient there is no information given. In our paper, when we compare the values of the coefficients for galaxy samples in narrow magnitude intervals with the values obtained from galaxy samples in wide magnitude intervals, we find that the maximum difference in the $a$ coefficient is approximately of the order 33\%, for the $b$ coefficient it is approximately 10\% and for the $c$ coefficient it is 20\%. Given the similarity between the differences in the coefficients in our work and those obtained using the data from D'Onofrio {\it et al}. (2006), we assert that an important reason for the differences found among the various authors may be caused by the different magnitude ranges within which the galaxies are contained, on top of the differences produced by the fitting method utilised and the independent variable used in this fitting method. On the other hand given that the differences found for the values of the coefficients $a$ and $b$ in our work are similar to those reported in the literature, we can safely say that the differences for the values of the $c$ coefficient from the literature will also be similar to the differences in the $c$ values reported here, approximately 20\%.

An important application of the FP is that it may be used to perform large scale distance estimates, however, the differences between the distances calculated utilising the values of the FP coefficients from different authors reported in D'Onofrio {\it et al}. (2006), may be significant. For example, if we assume that the FP is universal, that is to say, that the values of the coefficients $a$ and $b$ are consistent with a unique value, then the distances would be given as function of the differences in the zero point (see Blakeslee {\it et al}. 2002), then, if there is a 10\% error in the zero point this would produce a 50\% error in the distance determination. This is the error which could be made if the data reported in D'Onofrio {\it et al}. (2006) were used and it could also be the error made if the dependence on the magnitude range were not considered. Here we must stress that these distance uncertainties have been obtained without considering in the calculation the errors in coefficients $a$ and $b$, so the total error could be larger than 50\% both if we use the FP data from the different authors reported by D'Onofrio {\it et al}. (2006) as well as if we do not take into consideration the dependence of the FP coefficients on the magnitude range.

Finally, due to the change of the traditional FP coefficients with respect to the width and brightness of the magnitude range, it 
is important to consider that when comparisons between different ETGs samples are performed, it must be clearly established what is the magnitude range within which the ETGs are distributed. If this is 
not done the differences which might be found may be misinterpreted.

\section{Conclusions}

In this paper we make a compilation of photometric and spectroscopic parameters for 3 samples of ETGs from the literature, which include a total of approximately 8800 galaxies and cover a relatively ample 
magnitude range ($<\Delta M>$ $\sim 5.5$ $mag$). With this information we perform an analysis of the behaviour of the coefficients of the FP with respect to the width and brightness of the magnitude range. The results we obtain are as follows:

\begin{itemize}

\item We find that when a growing number of brighter galaxies is included in the calculations of the intrinsic dispersion and of the coefficients of FP (increasing magnitude intervals) or if we consider galaxy samples in progressively brighter fixed-width magnitude intervals (narrow magnitude intervals), their values change 
and these changes are larger than the associated errors for most of the cases.

\item This fact made us think that there might be a dependence both of the values of the coefficients of the FP relation and of the intrinsic dispersion of this relation with the width and brightness of the magnitude range. The existence of this dependence is confirmed with the application of a non-parametric test to the data (run-test).

\item Since the coefficients of the FP relation depend on the width and brightness of the magnitude range, we establish that this relation could be written as follows:

\begin{equation}
\log\,(r_{e})\, \; =\, \; a'\,\log\,(\sigma_{0})\, \; +\; b'\, \, <\mu>_{e}\;
+\; c'
\end{equation}
      

According to this, the traditional FP relation (equation 1) will produce the real coefficients only when we consider complete samples, as far as magnitude is concerned.

\item We find that the behaviour 
of the FP relation described above is similar to 
the bahaviour of the KR reported in Nigoche-Netro (2007) and Nigoche-Netro {\it et al}. (2007, 2008), that is, the dependence of the FP on the width and brightness of the magnitude range is caused by a geometrical effect due to the fact that the distribution of the galaxies in the $\log (r_{e}), <\mu>_{e}, \log(\sigma)$ space depends on luminosity. 

\item We find that any systematic restrictions imposed on a sample of ETGs, such as brightness cuts, effective radius cuts or velocity dispersion cuts will cause changes in the geometric form of the distribution of ETGs on the $\log (r_{e}), <\mu>_{e}, \log(\sigma)$ space. The more pronounced these changes are made, the more pronounced will be the changes in the values of the FP coefficients. 

\item Finally, due to the change of the FP coefficients with respect to the width and brightness of the magnitude range, it is important to remember that when comparisons between galaxy samples are performed, these must be done only for equivalent magnitude intervals, otherwise the results may be misinterpreted.

\end{itemize}

\section*{Acknowledgments}

We would like to thank Consejo Nacional de Ciencia y Tecnolog\'{\i}a (M\'{e}xico)
for a PhD fellowship number 132526, Ministerio Espa\~nol de Educaci\'on y Ciencia for grant PNAYA2006, Instituto de Matem\'aticas
y F\'isica Fundamental (CSIC, Espa\~na) and Instituto de Astronom\'ia (UNAM, M\'exico) for all the facilities provided
for the realisation of this project. We would also like to record our obligation to Prof. Mariano Moles for help with the 
presentation of this paper. Last but not least, we would like to acknowledge the many suggestions provided by an anonymous referee which greatly improved the presentation of this paper.

\begin{appendix}

\section{Analysis of the robustness of the run test using the bootstrap methodology}

\subsection{Run test robustness analysis for a set of general data}

We have run a number of experiments with the run test. This is the test which we use to establish the presence of an underlying trend in the behaviour of the coefficients of the FP. The experiments consist in:
\begin{itemize}

\item Generating a list of 200 random numbers between 0 and 1. A test of their distribution with the run test reveals that there is absolutely no underlying trend.

\item Following the Bootstrap method, we obtain 1000 resamples from the original sample and test their distribution with the run test. The character of their distributions changes from not having an underlying trend to having one in less than 20 cases in 1000 at a confidence level of 99\%.

\item Artificially introducing a monotonic trend.

\item Testing the new distributions with the run test reveals, in all the cases, the presence of an underlying trend at a confidence level of 99\%.

\item Now we introduce randomly an error which may be positive or negative and we test the distributions again. The errors introduced were $\pm$0.1, $\pm$0.2, $\pm$0.3, $\pm$0.4 and $\pm$0.5 or errors of the order 10\%, 20\%, 30\%, 40\% and 50\%.

\item For all the cases except for that with 50\% error, the run test continued to indicate the presence of an underlying trend at a confidence level of 99\%.

\item Again following the Bootstrap method, we obtain 1000 resamples from the original sample with random errors.

\item The run test reveals that in these thousand resamples with errors, the character of the distributions changes from having an underlying trend to not having one in less than 20 cases out of 1000 at a confidence level of 99\%.

\item The run test reveals that if the number of data for the original sample decreases and the error in the data increases then the number of resamples that change their distribution from having an underlying trend to not having one grows and the level of confidence at which we may reject the null hypothesis becomes smaller.

We conclude from this analysis that the run test predicts reliably the presence of an underlying trend in a set of data, even if these data have errors. Only when the errors are very large (\textgreater50\%), and the amount of data is small the run test is unable, in some cases, to detect the underlying trend present in the data.


\end{itemize}

\subsection{Run test robustness analysis for the FP data}

We ran a series of experiments on the values of the coefficients of the FP using the bootstrap methodology. One thousand samples were extracted from the original data (see Table 2). We also added to the data Gaussian errors. The maximum allowed error for each datum corresponds to the error for the real data. In Table A1 we show the results of the run test where the percentage given refers to the fraction of extracted samples in which an underlying trend has been detected at a level of confidence of 90\%. 

As can be seen in Table A1, when the maximum allowed error in the data of the extracted samples is of the same size as those errors affecting the real samples, we find that on the average 91\% of the extracted samples show an underlying trend at a confidence level of 90\%. We may again affirm that the run test is efficient in finding underlying trends even when the errors in the data have a significant value.

\begin{table}

 \centering
 \begin{minipage}{62mm}

\caption{Run test for the evaluation of the FP coefficients (increasing magnitude intervals) from the different samples of galaxies. One thousand samples were extracted from the original data following the bootstrap methodology. In addition the data were combined in a random way with Gaussian errors for which the maximum errors corresponded to the values cited in Table 2. The null hypothesis establishes that there is no underlying trend in the data. The percentages refer to the fraction of extracted samples for which an underlying trend is detected at a confidence level of 90\%.}

  \begin{tabular}{@{}lccccr@{}}

\hline

Sample   & $a$ & $b$ & $c$ \\

\hline

Total SDSS (g$^{\ast}$ filter) & 94\% & 94\% & 96\%\\
Total SDSS (r$^{\ast}$ filter) & 94\% & 93\% & 96\%\\
Total SDSS (i$^{\ast}$ filter) & 95\% & 93\% & 95\%\\
Total SDSS (z$^{\ast}$ filter) & 95\% & 95\% & 95\%\\
Coma (Gunn r filter) & 82\% & 82\% & 80\%\\
Hydra ( Gunn r filter) & 82\% & 95\% & 80\%\\

\hline

\end{tabular}
\end{minipage}

\end{table}

\end{appendix}

\bsp

\label{lastpage}

\end{document}